# An agent-based simulation model of pedestrian evacuation based on Bayesian Nash Equilibrium

Yiyu Wang[1], Jiaqi Ge[1], Alexis Comber[1]
gyywa@leeds.ac.uk; j.ge@leeds.ac.uk; a.comber@leeds.ac.uk

[1] School of Geography, University of Leeds, UK

**Abstract:**
This research incorporates Bayesian game theory into pedestrian evacuation in an agent-based model. Three pedestrian behaviours were compared: Random Follow, Shortest Route and Bayesian Nash Equilibrium (BNE), as well as combinations of these. The results showed that BNE pedestrians were able to evacuate more quickly as they predict congestion levels in their next step and adjust their directions to avoid congestion, closely matching the behaviours of evacuating pedestrians in reality. A series of simulation experiments were conducted to evaluate whether and how BNE affects pedestrian evacuation procedures. The results showed that: 1) BNE has a large impact on reducing evacuation time; 2) BNE pedestrians displayed more intelligent and efficient evacuating behaviours; 3) As the proportion of BNE users rises, average evacuation time decreases, and average comfort level increases. A detailed description of the model and relevant experimental results is provided in this paper. Several limitations as well as further works are also identified.

## 1 Introduction

Large public gatherings or crowds are commonplace and have been the subject of simulation research in many studies related to crowd management, disaster management and evacuation planning (Babojelić and Novacko 2020). However, in-depth research on pedestrians has been hindered by difficulties such as complex individual behaviours, different disaster characteristics, and varying environmental factors (Wijermans and Templeton 2022). As evacuee behaviour and movement vary in different scenarios, a number of field observations and simulation experiments have been conducted to explore pedestrian flows, movement patterns and potential factors affecting evacuation under different types of emergencies (Rozo et al. 2019; Feng et al. 2021; Sevtsuk and Kalvo 2022). Despite many simulation studies of pedestrian behaviours, few common behavioural features of pedestrian flows have been explored (Vermuyten et al. 2016; Babojelić and Novacko 2020). One of the main obstacles is the lack of experimental datasets that closely match individual movements during evacuations in the real world. Consequently, a more intelligent evacuation simulation model of pedestrian flow is needed to realistically replicate the movement and behaviours of evacuees and that is also easily adaptable to various evacuation scenarios.

Simulation models of pedestrian flow are generally classified into one of two main categories: macroscopic models and microscopic models. Macroscopic simulation models consider pedestrian flows as a single unit such that evacuees in these models are homogeneous during simulation (Jiang et al. 2010; Piccoli and Tosin 2011). In these it is difficult to observe the interactions of individual pedestrians and their (micro-level) behaviours. To address this, a number of approaches have been developed, such as cellular automata, lattice gas automata, social force models, and other simulation tools, allowing pedestrians to be partially heterogeneous during simulation. Pedestrians determine their own actions according to their surrounding environment, but the probability distributions of their decisions are still controlled at the macroscopic level (Teknomo 2016; Lu et al. 2017). Agent-based modelling (ABM) can fully capture individual behavioural heterogeneity in pedestrian models, and it is one of the main individual-based models used to simulate pedestrian movement in different scenarios. It has the capability to reveal the aggregated patterns of individual actions and environments from the bottom up (Bar-Haim 2010).

Game theory has attracted much attention from researchers in the fields of pedestrian behaviours and evacuation simulation. Current research on pedestrian flow pays much attention to whether and how the simulation could closely match the reactions of pedestrians in different scenarios in the real world by predicting pedestrians' next move. Game theory provides an effective approach to realistically reproduce individual decision-making processes. Specifically, a game-theoretic approach assumes individuals make decisions based on their beliefs, which are updated in response to their surroundings. The best pedestrian strategies or responses can be derived using different game theories. For instance, Rigos et al. (2019) introduced Sequential Equilibrium and perfect Bayesian Equilibrium into their response model to simulate the evacuee actions after receiving the order to evacuate. Liao et al. (2019) incorporated Bayesian Nash Equilibrium into their simulation model to discover the relationships between safe pedestrian flow rates and public space area. There are many other examples of research on pedestrian behaviours that have introduced game theory into their models to more realistically simulate pedestrian decision-making and behaviours (Bouzat and Kuperman 2014; Lo et al. 2006; Mesmer and Bloebaum 2016). However, the main objectives of the studies focusing on both game theory and ABMs have generally been to compare simulations based on game theory with agent-based approaches (Norri et al. 2021), or the incorporation of ABMs with simple game theory such as a zero-sum 2-player game (Lo et al. 2006; Levy et al. 2018). Research using Bayesian game theory for pedestrian evacuation have tended to simulate individual selection of final exit rather than their actions during evacuation process (Bouzat and Kuperman 2014; Mesmer and Bloebaum 2014). The focus of these simulations was the interaction between a small number of agents rather than the mutual influences of a large number with the environment. In summary, game theory has been widely adopted in the context of individual evacuation decisions such as exit selection and route optimization (Levy et al. 2018; Mesmer and Bloebaum 2014) and is regarded as an appropriate behavioural model to simulate pedestrians' actions under emergencies.

Few studies have sought to simulate pedestrian behaviours during emergency evacuation using game-theoretic approaches. One of the main barriers is that the interactions between individual behaviours and macro-phenomena of pedestrian flow are complex, and general game theories cannot account for the complexity of interactions between and among pedestrians, as well as with their environment. Early game theories place a number of restrictions on agents and environments such as the Nash game with complete information (Rosen 1965) and team-based decision making (Radner 1962). The refined Bayesian Nash Equilibrium (BNE) proposed by Ui (2016) relaxes the complete information constraint and considers a game with incomplete information, which is more realistic in the context of evacuation when complete real-time information is often missing for individuals. BNE defines a correlated equilibrium with varied payoff gradients according to different game conditions, which include the Nash Equilibrium as one particular case, compared to the monotonic payoff gradient in traditional Nash Equilibrium. As a result, BNE is more suitable to scenarios with incomplete information, multiple equilibriums, and varied payoff gradients, suggesting opportunities to use it to simulate pedestrian movement in an ABM. BNE has been used mainly in advertising and other economics fields (Gomes and Sweeney 2014) with little research using BNE to simulate pedestrian evacuation.

This research develops an agent-based evacuation simulation model of pedestrian flow by introducing BNE to realistically simulate individual decision-making processes and pedestrian behaviours in an emergency evacuation. The model combines Bayesian game theory and an agent-based approach at an individual level, to provide an experimental environment of pedestrian flows to support research on crowd management and evacuation planning. It was hypothesized that a BNE approach was able to positively affect individual evacuations during model simulation because of its capacity to predict future congestion levels. A series of simulation experiments were conducted with different parameter configurations to understand whether and how BNE affects pedestrian emergency evacuations.

# 2 Model Description

A complete and detailed description of the initial model following ODD+D protocol (Müller et al. 2013) is provided in this section.

## 2.1 Overview

### 2.1.1 Purpose

The purpose of this model is to introduce a new individual decision-making method, BNE, into the ABM of pedestrian evacuation to simulate individual behaviours and movements. The model was built to balance between fast evacuation and high comfortability, which is a general conflict in the domain of pedestrian research. The interactions of pedestrians with their neighbours as well as surroundings was also considered in order to simulate a more realistic pedestrian evacuation. This model ultimately aims to explore the influences of BNE on pedestrian flows from various perspectives, especially pedestrian comfort and exit time in an emergency evacuation with different parameter configurations.

### 2.1.2 Entities, state variables and scales

The model contains two main types of entities: Patches (i.e. evacuation space) and Agents representing evacuating pedestrians. The variable names are same as the variables implemented in NetLogo.

The **Global Environment** is defined as model parameters at the system level, controlling all of the global variables representing the simulation environment. Its state variables are shown in Table 1.

Table 1. Global environment state variables

| Variable Name | Variable Type and Units | Brief Description |
| --- | --- | --- |
| **number-persons** | Person | Total number of agents in the model |
| **Percentage-of-agents-with-BNE** | Percent | The proportion of agents who are using BNE to evacuate |
| **Probability-competing** | Percent | The probability of agents entering one patch |
| **door-width** | Patch | The size of exits |
| **move-speed** | m/s | The speed when agents can move freely which reduces with the increasing crowd density |
| **Step-length** | m | The length of a single agent step |
| **follow-radius** | Patch | The distance that agents could see; it is used in Random Follow patterns |
| **weight-Ud** | Numeric | A coefficient to balance the influence of distance utility and expected comfort utility on determining agent movement directions. |
| **Moving-pattern** | Chooser | 4 patterns are available: Shortest Route (SR), Random Follow (RF), BNE mixed with SR, and BNE mixed with RF |

**Patches** refer to the areas in the simulation space. The evacuation environment in this model was divided into 1360 (68*20) patches and as values of different utilities can control the directions of agents, these patch attributes are considered as state variables. Details of the patch state variables are shown in Table 2.

Table 2. Patch state variables

| Variable Name | Variable Type and Units | Brief Description |
|---|---|---|
| **Uec** | Numeric | Expected Comfort Utility |
| **Ud_lt** | Static; Numeric | Distance Utility, used by the agents moving to the left exit |
| **Ud_rt** | Static; Numeric | Distance Utility, used by the agents moving to the right exit |
| **U_total** | Numeric | Total Utility, the sum of distance utility and expected comfort utility |
| **patch-target** | Patch | Patches with maximum total utility |

**Agents** represents the individual evacuees with different movement patterns and the related state variables are shown in Table 3.

Table 3. Agent state variables

| Variable Name | Variable Type and Units | Brief Description |
|---|---|---|
| **speed** | m/s | The speed of each agent during evacuation |
| **left?** | True/False | Whether or not the agent moves to the left exit |
| **follow?** | True/False | Whether or not the agent follows another agents |
| **door** | Location | The location of exit that the agent chooses |
| **BNE-type** | Boolean | "1" – this agent uses BNE to evacuate; <br> "0" – agent follows SR/RF patterns |
| **nearby-leaders** | Agentset | The optional neighbours when the agent want to choose one to follow; only used in RF patterns |
| **leader** | Agent | The neighbour followed by the agent |

**Scales.** The spatial extent of this model is a rectangular region of 68 * 20 square patches (see Fig. 1). The model space is bounded and agents can only evacuate through the exits on either side. The model runs until all the agents evacuate from the simulation space. That is, the temporal scale in this model was not absolute and the number of time steps depended on the initial environmental conditions and the agents themselves. Three behavioural models were evaluated: Shortest Route (SR), Random Follow (RF) and BNE. The behavioural models were used to generate four moving patterns (i.e. model configurations): SR, RF, BNE mixed with SR, and BNE mixed with RF. The moving pattern is selected by the user at the beginning of the simulation.

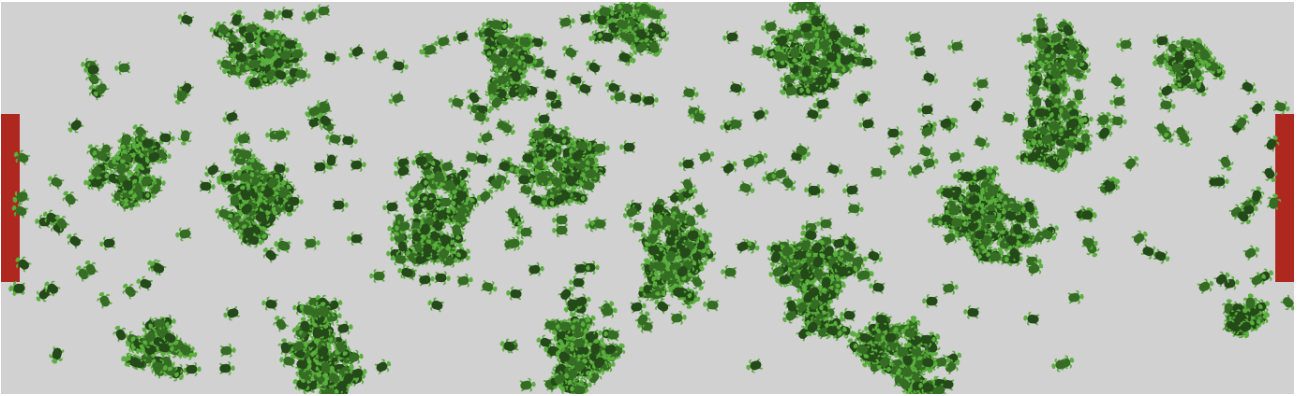
Figure 1. The interface of the simulation model, with agents in green and exits in red.

**2.1.3 Progress Overview and Scheduling**
The model simulates the complete process of pedestrian evacuation and demonstrates the detailed decision-making process of agents especially using BNE. Over each simulation run, patches and agents continuously update the relevant state variables in each time step.

The schedule of the model is shown as follows:
I. The simulation begins with a series of initial settings for global environment by the user, including the percentage of BNE agents, exit size, the number of agents, and other state variables. The type of moving pattern is also selected, with all the environmental attributes static until the end of this run.
II. The patches execute the calculations of their distance utility and expected utility which are related to the implementation of BNE in the model. Expected comfort utility is continually updated until the end of simulation.
III. The agents choose a new direction (i.e. repeat their decision-making process) every time step in response to the new environmental conditions.
IV. The state variables, plots and model interface are updated.

## 2.2 Design Concept

**2.2.1 Theoretical Background**
Bayesian Nash Equilibrium (BNE) was used in the individual decision-making process of this model in order to augment the rationality of pedestrian evacuation simulation. BNE is an extension of Nash equilibrium and is a static game with incomplete information. It is generally defined as a strategy profile in which participants are assumed able to maximize their own expected utility based on the probability distributions of the strategies chosen by other players, and in this case, no player selects other strategies (Ui 2016). That is, the primary element of individual decision-making is the probability distributions of strategies played by other nearby participants, especially the probabilities of neighbours choosing the same strategy as the player. In this model, this is reflected as the probability distributions of the next actions of nearby agents and a series of utility calculations for candidate patches. The relevant underlying equations are described below.

**2.2.2 Individual Decision-making**
Decision-making is modelled on an agent level in this research. In this model, the decision-making processes of agents are different based on the three behavioural models (i.e. SR, RF and BNE). For BNE agents, they tend to move to a neighbour patch with maximum total utility every time step, which depends on the probabilities of nearby agents' next actions and the surrounding environment. Each agent using RF follows a neighbour in sight at random and repeats this selection process until the end of simulation. Agents using SR makes only one decision (i.e. which exit to move towards) at the beginning of the simulation, which means that their directions remain unchanged until they succeed in evacuating from the simulation space.

### 2.2.3 Individual Sensing
Agents following the RF pattern are assumed able to sense their neighbours within a specific radius of their current locations. The value of this radius remains constant during simulation and can be adjusted by the corresponding slider "follow-radius". Agents followed BNE pattern are assumed to be able to sense surrounding conditions and select the target from nearby patches to move in each time step. Specifically, an agent using BNE can sense and potentially move to all neighbouring patches which are located between its current patch and the exit. The choice criteria are the total value of distance utility and expected comfort utility of each patch.

### 2.2.4 Individual Prediction
Individuals in this model use the information on neighbours' current locations and the probability distribution of their moving directions to predict future situations. Specifically, expected comfort utility of neighbouring patches is estimated over a time step by using the explicit prediction of nearby agents *p(n)* and comfort coefficient of each patch $U_c(n)$.

### 2.2.5 Interaction
Interactions among agents are mediated through the variations of expected comfort utilities of neighbour patches. In BNE patterns, each patch's expected comfort utility depends on the expected number of agents who will be on the patch in the next step, which in turn determines the number of agents in the nearby patches. Then, each agent will determine its direction by comparing the total utilities of six patches which are within its optional directions (i.e. candidate patches $P_0$, $P_1$, ..., $P_5$) (details in Section 2.3.4). That is, the current position and expected next move of agents influence the expected comfort utility of patches, which, in turn, affects the next move of nearby agents.

### 2.2.6 Heterogeneity
Agents are heterogeneous in their decision-making process based on their own behavioural type. For agents following the Shortest Route pattern, their only decision is to move to the exit. For those with the Random Follow pattern, agents need to choose a neighbour in their view each time step and repeat this process until the end of simulation. For those with BNE pattern, agents select the patch with maximum total utility to move to and repeat this decision-making procedure every time step until they evacuate successfully.

### 2.2.7 Stochasticity
Stochasticity is introduced in two ways in this model. Firstly, the model is initialized randomly based on the settings configured by the user. Specifically, (a) the location of agents, (b) the random allocation of behavioural type, and (c) the initial directions of agents are considered to be stochastic at the beginning of a simulation. Secondly, when an agent following the RF pattern determines where to move, its choice of its following target is partly stochastic as it is limited by its view and the exit selected. Similarly, when two patches have same and maximum value of total utility, the agent in BNE pattern will randomly select one to move to. This decision is stochastic but not completely random since the choice is restricted by the location and direction of agents.

### 2.2.8 Observation
The purpose of this model is to explore whether and how BNE affects pedestrian evacuation procedures in the case of emergency, with two main measurements: evacuation time and pedestrian comfort level. The exit time and average expected comfort utility of each run are collected at the end of simulation in order to compare evacuations with varying proportions of agents following the BNE pattern in the simulation.

## 2.3 Details

### 2.3.1 Implementation Details
The initial model was developed in NetLogo. The source code and experimental data are available at https://www.comses.net/codebases/fd29c3e3-c4b4-4360-b13a-4e05ce287b02/releases/1.0.0/ .

### 2.3.2 Initialization
The initial state of the model is a hypothetical evacuation space with emergency exits located on either side. The agents are initialized by setting the total number of persons (i.e. number-persons global parameter) and the percentage of BNE users through a slider Percentage-of-agents-with-BNE. The agents are randomly scattered over the simulation environment. The initial speed of each agent is tailored according to the number of agents on its nine neighbouring patches, and all the adjustments are based on the reference speed assigned by the global parameter move-speed. The agent moving patterns are selected through the moving-pattern. The moving patterns in this model consist of Random Follow (RF), Shortest Route (SR), BNE mixed with RF, and BNE mixed with SR. For the first two patterns, all the agents are set to same moving pattern during simulation. For the last two patterns, a specific proportion of agents use BNE to evacuate which is defined by the global parameter Percentage-of-agents-with-BNE, and the rest follow one of two other patterns (i.e. SR or RF) based on the selected option.

Each patch calculates its own distance utility and expected comfort utility at the initialization stage. BNE agents compare the value of total utility (i.e. the sum of $U_d$ and $U_{ec}$) for the candidate patches and select the one with maximum value to move to in each time period. That is, the patch attributes are being continuously updating every time step to provide updated information to agents using BNE for determining their next actions. At present, a series of global parameters (e.g. door-width, follow-radius, etc.) are fixed due to the main research focus, which is the exploration of whether and how BNE affects pedestrian emergency evacuation, but variations in these global parameters could be evaluated in further research.

### 2.3.3 Input data
So far, no input is read in this initial model.

### 2.3.4 BNE
In order to translate the rationality of BNE theory into specific decision-making rules, a series of utility functions are introduced in this model to realize the BNE behavioural model. Individual decision-making depends on the value of "Total Utility" for optional patches. Total utility consists of three main elements: Distance Utility ($U_d$), Comfort Utility ($U_c$) and Expected Comfort Utility ($U_{ec}$), and refers to the total value of $U_d$ and $U_{ec}$. Specifically, the decision made by each BNE agent considers the distance from its current position to the exit, the number of neighbours who may move to the same patch as itself, and the possible surrounding situations in the next time step. Then, the patch with maximum total utility (i.e. the sum of $U_{ec}$ and $U_d$) is selected by the agent to move to. In other words, agents use BNE to predict the congestion level in next time step and then avoid the most clogged patches during their movement, in order to determine an evacuation route with less exit time and higher comfort level. The choice criterium is the value of total utility in neighbouring patches, which is evaluated by agents to decide where to go. In this model, all BNE related utilities were set as patch attributes and are described in detail below.

*A. Distance Utility.*
This represents the distance from the current location to the exit. Since we assume that agents tend to choose the patch with largest value of total utility, $U_d$ should be set to an increasing attribute value closer to the exits. Due to two exits existing in the evacuation space, two sets of distance utility are determined for the agents moving to the right or left exit respectively (i.e. parameters $U_{d\_rt}$ and $U_{d\_lf}$). The equation is:

$$U_d = \frac{D-d}{D} \qquad (1)$$

where, *d* represents the distance from current patch to the exit, and *D* refers to the diagonal of the evacuation space.

*B. Comfort Utility.*
Comfort Utility, $U_c$ is a set of coefficients that form a crucial component of Expected Comfort Utility, reflecting the comfort level of agents in any one patch. According to the speed-density relation associated with the Spatial-Grid Evacuation Model (SGEM) (Lo et al. 2004), the value is set to 1 when two or less than two agents occupy the patch. It decreases as the number of agents on the patch increases, by setting the value to be a proportion of the free-moving speed (i.e. 1.4 m/s) relative to the number of agents on the patch. Considering the limited space capacity in reality, $U_c$ stays at zero when more than 4 persons move to the same patch. The equation is as follows, with full details in the calibration section (Section 3.1):

$$U_c(n) = \begin{cases} 1.00, n \leq 2 \\ 0.51, n = 3 \\ 0.07, n = 4 \\ 0.00, n \geq 5 \end{cases} \qquad (2)$$

where, *n* represents the number of agents in one patch.

*C. Expected Comfort Utility.*
According to the definition of BNE, individual decision-making in this model is independent, which means that no account is taken of the agent's previous actions in each time step. The main factors determining where agents go is the number of agents who may move to the candidate patches in next time step. In other words, the probability of the neighbours' next actions has an impact on the decision-making process of the agent.

It is assumed that the probability of reverse movement during evacuation is extremely low, which means that each agent has six optional directions (i.e. candidate patches) $P_0, P_1, …, P_5$ in each time step (see Fig. 2). The probability of entering these candidate patches Pm is set to the same value (i.e. 16.7%) by default, which could be adjusted using the Probability-competing slider in further studies.

Thus, the patch variable named Expected Comfort Utility ($U_{ec}$) for each patch is dynamic in this model and reflects the interaction between agents. It is defined as the multiplication of comfort utility $U_c$ and the probability *p(n)* that a certain number of agents move to this patch in next time step (see Eq(3)):

$$U_{ec} = \sum_{n=0}^{4} p(n) U_c(n)$$
$$= \sum_{n=0}^{4} C_N^n P_m^n (1-P_m)^{N-n} U_c(n) \qquad (3)$$

where, *n* represents the number of agents in this patch; and $P_m$ refers to the probability of agents entering the candidate patches, which is set to 16.7% by default. In this way, the calculation of $U_{ec}$ takes into account the agents on both the patch and its nine neighbouring patches.

The relationships of these utilities are illustrated as Fig. 3.

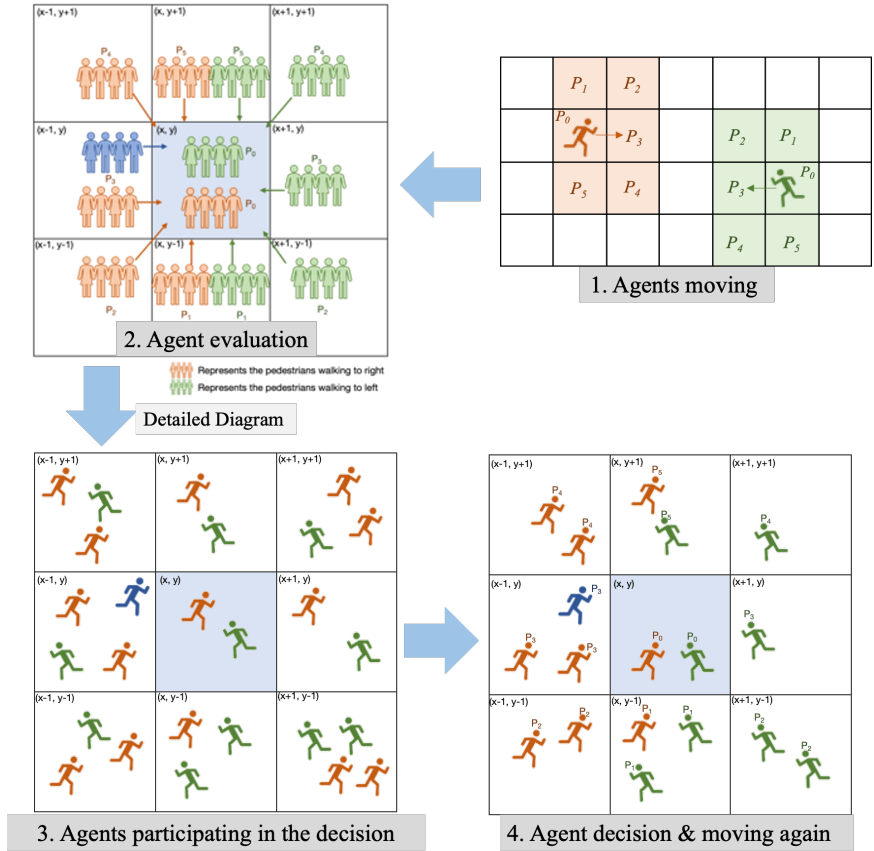

Figure 2. The scheme of agents' decision-making

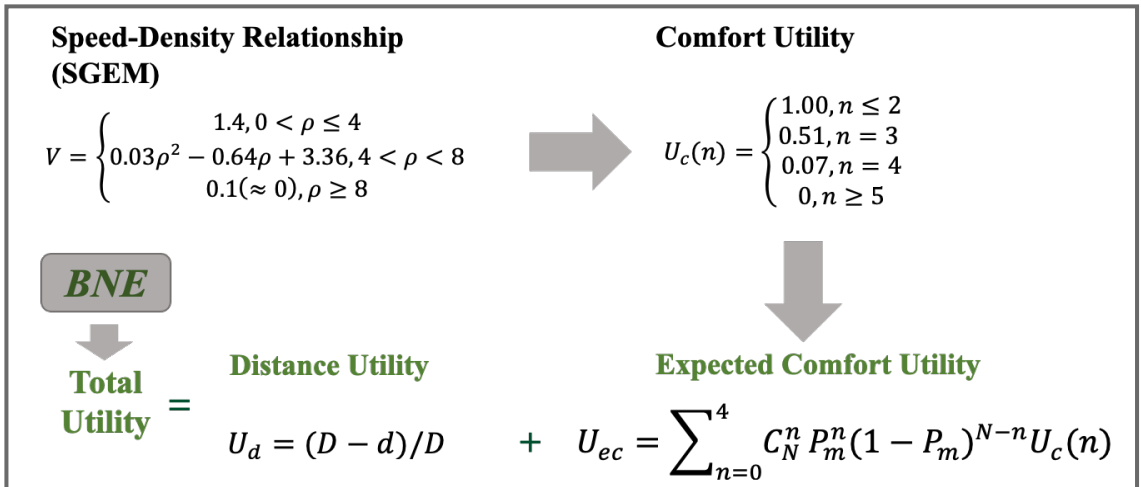

Figure 3. The schema of BNE utilities

# 3 Calibration, Simulation Experiments

## 3.1 Speed Calibration

To achieve a more realistic evacuation simulation, it is assumed that individual speeds in the model should change over time instead of a static attribute. The variation of moving speed has a close association with the number of surrounding agents, which means that the speed parameter should be calibrated in this way. After the comparison of several of the main pedestrian speed-crowd density models adopted in recent years (Mesmer and Bloebaum 2016; Luo et al. 2018; Zhou et al. 2019), the Spatial-Grid Evacuation Model (SGEM) proposed by Lo et al. (2004) was considered as the most appropriate relation model for this research, as it takes into account of the interconnections between surrounding pedestrians, as well as the potential effects of the short-term contact among pedestrians on individual evacuation speeds.

The general trend in the speed-density relation model remains consistent when the crowd density is less than 4 person/m$^2$, with pedestrians in a free motion state with a speed of about 1.4 m/s. However, when crowd density is greater than 8 person/m$^2$, pedestrians are considered to be in a state of constrained motion and move at around 0.1 m/s. When crowd density varies between this range between 4 person/m$^2$ and 8 person/m$^2$, pedestrian movement starts to be restricted and movement speed declines with increasing numbers of people. As the average step size of adults is around 0.7m who have a mean response time of about 0.5s (Chang et al. 2021), several parameters are adjusted in the SGEM model to fit the current environment. In this case, the initial speed in the model is adjustable through the move-speed slider instead of imposing a fixed value (i.e. 1.4 m/s), and individual speed is set to be in an inverse proportion to the number of nearby agents. Consequently, the suitable speed-density relationship for this model is illustrated in Eq(4):

$$V = \begin{cases} 1.4, 0 < \rho \leq 4 \\ 0.03\rho^2 - 0.64\rho + 3.36, 4 < \rho < 8 \\ 0.1(\approx 0), \rho \geq 8 \end{cases} \quad (4)$$

where, $\rho$ is density of pedestrians (person/m$^2$).

## 3.2 Simulation Experiments

A series of simulation experiments were conducted in NetLogo BehaviorSpace to evaluate the role of BNE played in pedestrian evacuation. To determine whether and how BNE affects individual evacuation, three simulation scenarios were evaluated (1) **Singleton Pattern with a fixed number of persons**: evacuations in which all pedestrians follow one of three moving patterns (i.e. SR, RF, BNE), (2) **Mixed Pattern with a fixed number of persons**: evacuations in which a specific proportion of agents follow BNE pattern and the rest evacuate by one of the other two patterns (i.e. BNE mixed with SR/RF), and (3) **Mixed Pattern with a varied number of persons**: evacuations in which the number of persons varies in a range from 1100 to 3000 and two types of agent movements participate into the simulation (i.e. BNE mixed with SR/RF). The evacuations were evaluated in terms of evacuation time and expected comfort utility to assess the performance of BNE.

The first experiments simulated evacuations in a tunnel space with all agents following one of BNE, RF, and SR patterns, with 2000 or 3000 pedestrians. The experiments were replicated 50 times for each parameter configuration and stopped when all of the agents evacuated successfully.

The second set of experiments evaluated how different proportions of agents following BNE would affect evacuation. The model was initialized with 2000 or 3000 agents, in which a varying proportion of agents followed BNE, and the rest SR or RF. Here the percentage of BNE users was set to vary from 0% to 100% in intervals of 2%. Simulations were replicated 50 times for each parameter configuration to evaluate the variations of both exit time and expected comfort utility.

The third set of experiments then simulated different numbers of pedestrians in the same tunnel space to explore the influence of BNE on evacuation scenarios with different crowd densities. Here the agents were randomly scattered over the simulation space with mixed moving patterns (i.e. BNE mixed with SR or BNE mixed with RF). The number of people varied from 1100 to 3000 in intervals of 100, and the proportion of BNE users was set to vary from 0% to 100% in increments of 2%. In this case, 30 repetitions were undertaken for each parameter configuration.

The values of all the other variables remain unchanged to in all experiments. The list of model inputs is shown in Table 4.

Table 4. The list of parameter settings in experiments

| Parameters | Values (Experiment 1) | Values (Experiment 2) | Values (Experiment 3) | State |
|---|---|---|---|---|
| number-persons | 2000/3000 | 2000/3000 | 1100 to 3000 (+100) | Dynamic |
| Percentage-of-agents-with-BNE | 100% | 0% to 100% (+2%) | 0% to 100% (+2%) | Dynamic |
| Probability-competing | 16.7% | 16.7% | 16.7% | Static |
| door-width | 6 | 6 | 6 | Static |
| move-speed | 2 m/s | 2 m/s | 2 m/s | Static |
| Step-length | 0.7 m | 0.7 m | 0.7 m | Static |
| follow-radius | 3 | 3 | 3 | Static |
| weight-Ud | 1 | 1 | 1 | Static |
| moving-pattern | Shortest Route; Random Follow; BNE mixed with SR/RF | BNE mixed with SR/RF | BNE mixed with SR/RF | Dynamic |

## 4 Results

To evaluate the effects of BNE on pedestrian evacuation, evacuation time and expected comfort utility were determined for each set of runs. Evacuation time is the time taken for all agents to successfully evacuate from the simulation space. Expected comfort utility refers to the overall mean of $U_{ec}$ recorded in each time step during a simulation.

For the first set of experiment, the evacuation time for each of three behavioural patterns (i.e. BNE, RF, SR) were evaluated for 2000 and 3000 persons (Fig. 4). The results show that BNE exit time was nearly half that of Random Follow and around two-thirds of the Shortest Route time. Similar findings were found in both two groups, demonstrating the impact that BNE has on reducing evacuation times compared with other more general moving patterns.

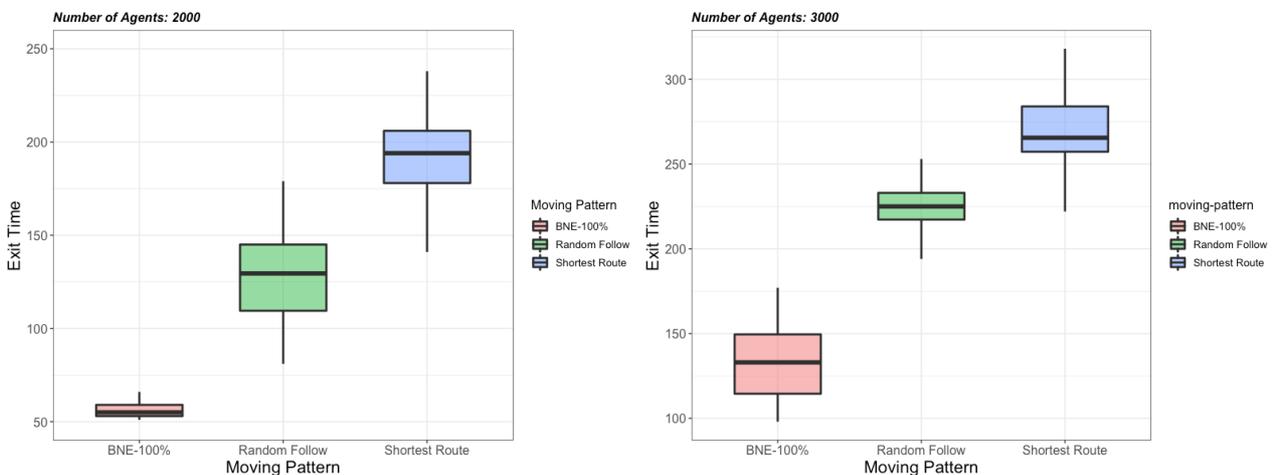

Figure 4. Evacuation time of 3 movement patterns: BNE, RF, SR (Number of Agents: 2000 and 3000).

Additionally, specific stages of the behavioural models were also recorded during simulation. The model views were exported every 20 ticks and the stages of models in the first 100 ticks are illustrated in Fig. 5. As shown, the congestion levels were more severe in the Shortest Route than the Random Follow, which were in turn more severe than BNE. This potentially explains the poorer performance of SR compared to RF. SR agents move forward to the exits all the time causing jams especially as more agents try to evacuate through the exits (which may not be wide enough). RF agents randomly follow one of their neighbours resulting in smaller scale of crowd groupings during evacuation. So, congestions are present in both the RF and SR patterns, but the relatively smaller scale of congestion in RF results in a quicker evacuation. BNE pedestrians were able to forecast the level of congestion in the next time step to avoid jams.

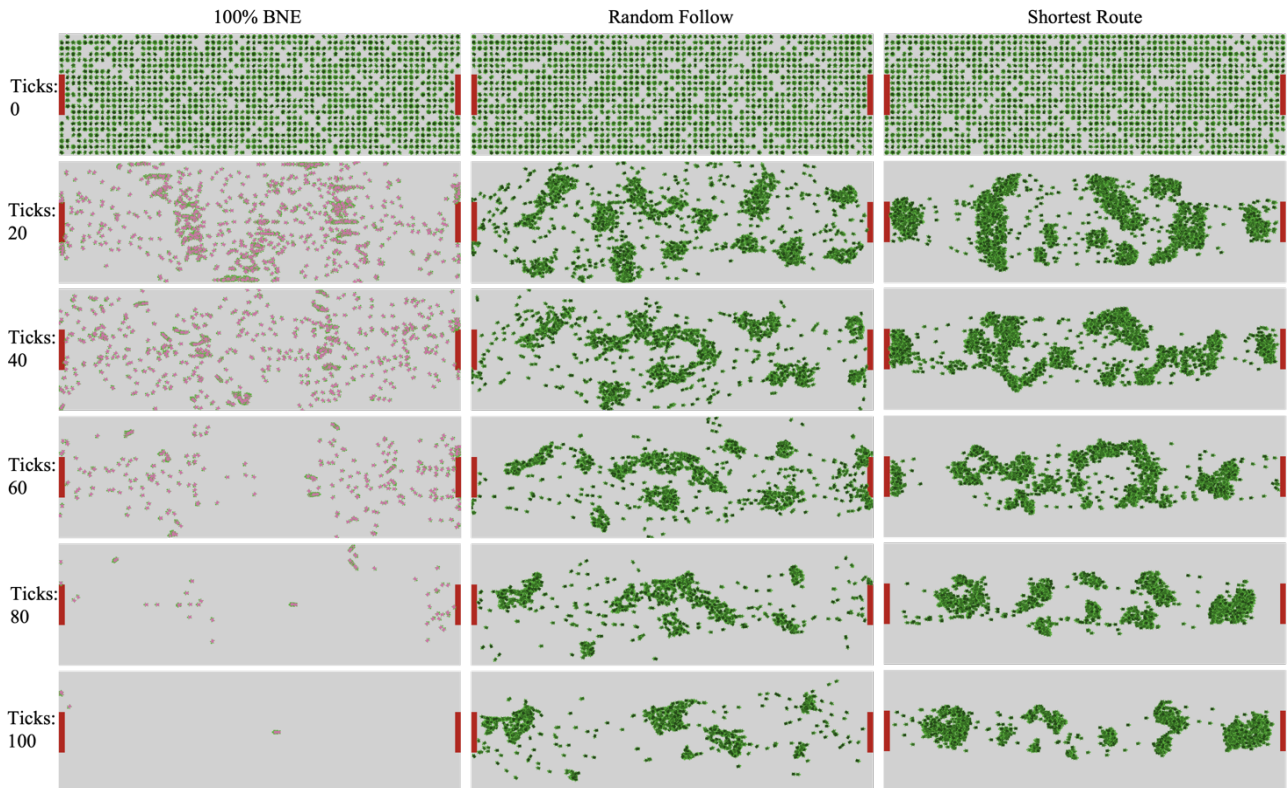

Figure 5. The stages of the flow of agents in three moving patterns -100%BNE, RF and SR

The next set of experiments evaluated expected comfort utility and exit times. BNE was combined with the other behavioural patterns (SR and RF) to understand how different proportions of BNE users affected evacuation. A series of simulations were conducted in which the percentage of agents following BNE varied from 0% to 100% at intervals of 2%.

The model was first simulated with 2000 persons and varying levels of BNE mixed with SR and BNE mixed with RF. Fig. 6 & 7 illustrate the variations in exit time and mean expected comfort utility of these two patterns respectively. As shown in Fig.6, evacuation time decreases as the number of BNE users increases in both two experiments and flatten at about 80% BNE users. Fig. 7 shows the variations of average expected comfort utility in two experiments. Higher expected comfort utility values indicate that evacuating pedestrians feel more comfortable during simulation. Fig. 7 illustrates a decreasing trend of $U_{ec}$ in BNE mixed with RF and an increasing one for BNE mixed with SR with a peak around 40% agents using BNE and 60% with the SR pattern. That is, BNE shows a significant positive effect on improving individual comfort level when we mixed BNE and SR patterns.

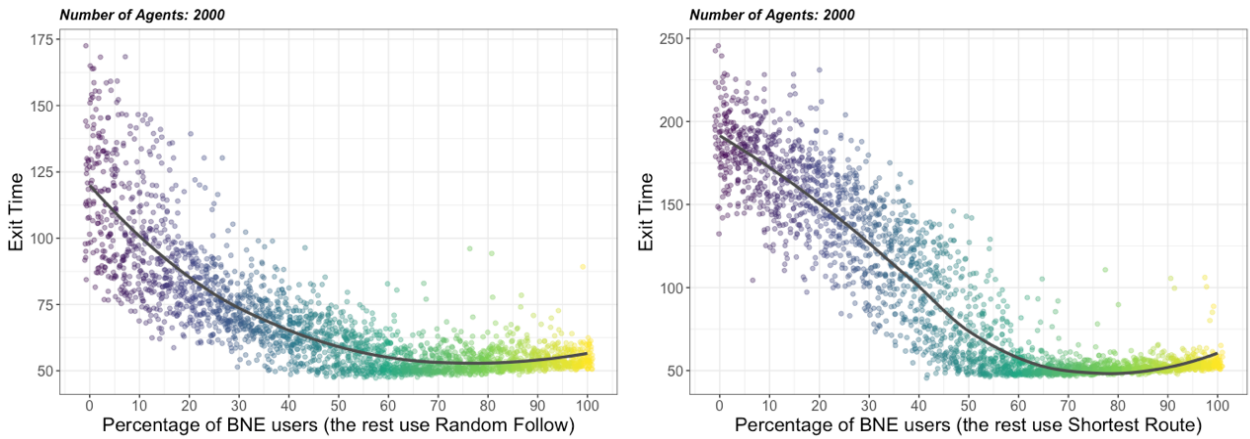
Figure 6. The variations in evacuation time of the patterns: Mixed Patterns (2000 persons)

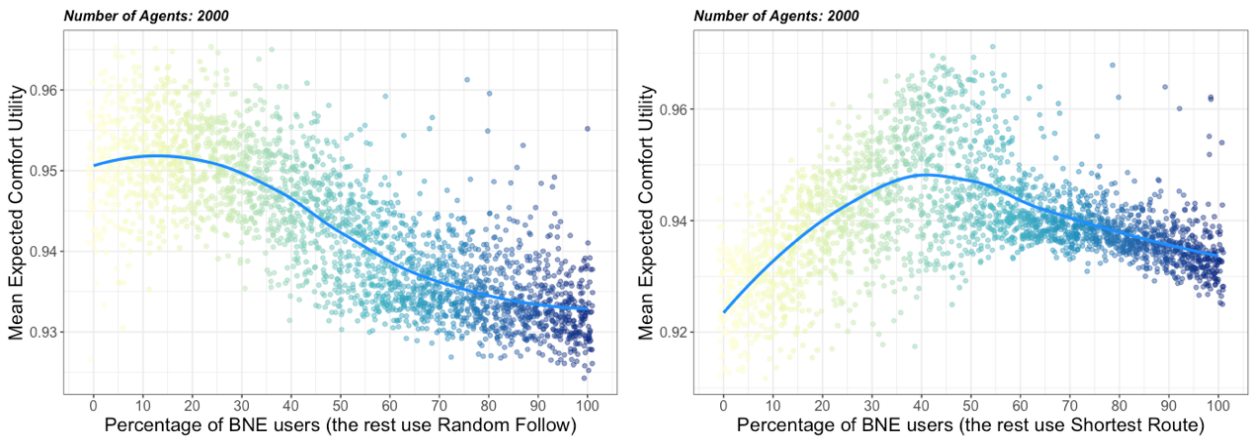
Figure 7. The variations in average expected comfort utility of the patterns: Mixed Patterns (2000 persons)

To further confirm the influence of BNE, the number of agents was set 3000 and the experiment repeated. Fig.8 shows a strong downtrend in evacuation time with increasing proportions of BNE and upward trends after around 60%-70% with RF and 70%-80% with SR. The average $U_{ec}$ showed an increasing trend with BNE with both RF and SR (Fig. 9), indicating the positive effects of BNE on reducing evacuation time, as well as improving individual comfort during evacuations.

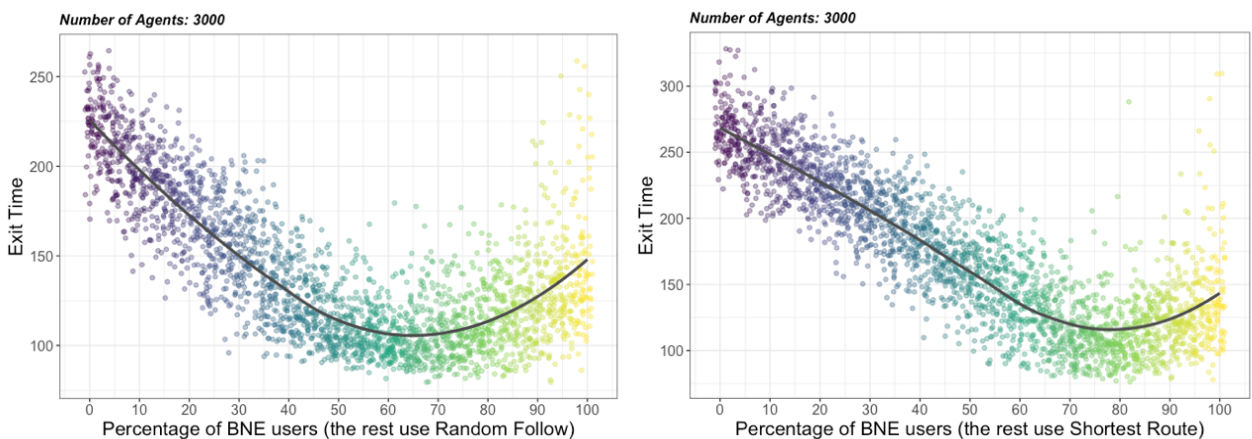
Figure 8. The variations in evacuation time of the patterns: Mixed Patterns (3000 persons)

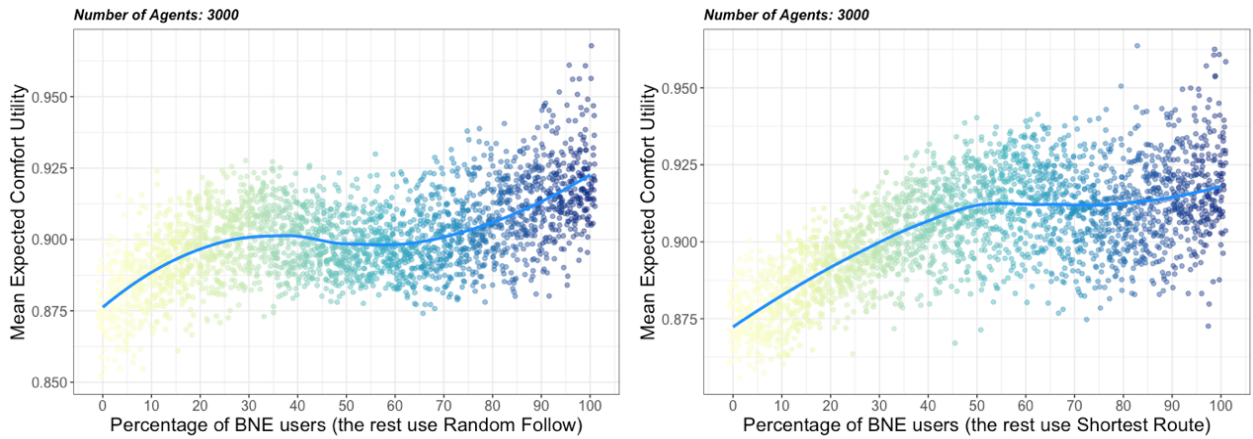
Figure 9. The variations in average expected comfort utility of the patterns: Mixed Patterns (3000 persons)

A potential reason for the contradictory trends of average $U_{ec}$ in BNE mixed with RF pattern (Fig. 7&9) may be the difference in crowd density. With lower density (i.e. 2000 persons), BNE has a notable influence on improving pedestrian comfort. To unpick this further a third set of experiments were conducted with the number of agents varying from 1100 to 3000 in increments of 100, and both exit time and average $U_{ec}$ were recorded to further evaluate the effects of BNE on pedestrian evacuation with different crowd densities. The full results are shown in Appendix A & B. Combining BNE with RF pattern (Appendix A), the advantage of BNE is not found evacuations with low crowd density. However, a distinct decrease of evacuation time with increasing BNE proportion can be observed in the scenarios with over 1500 agents. Thus, the positive influence of BNE on reducing exit time becomes increasingly evident with increased crowd density. Similar features are also appeared in BNE mixed with SR (Appendix B): exit time declines with increasing BNE proportions as the densities increase, but with a greater positive effect at lower crowd densities than in BNE with RF.

In addition, the variations of average $U_{ec}$ with the increasing number of agents in BNE mixed with RF/SR were also demonstrated in Appendix C&D respectively. Since BNE was perceived to have no influences on improving pedestrians' comfort when the number of agents was set to 2000 in BNE-RF pattern (Fig. 7), the range of initial pedestrian numbers was extended and an upward tendency of average $U_{ec}$ when the number was set to 2300 and was increasingly evident with increased agent density (Appendix C). As shown in Appendix D, uptrends were observed in all scenarios under mixed BNE with SR and peaked at around 50% (BNE users) in most of the plots. Average $U_{ec}$ increases with the growing proportion of BNE users in cases with high crowd density (e.g. 3000 persons).

Therefore, BNE was shown to have significant beneficial influences on shortening evacuation time as well as improving pedestrian comfort during emergency evacuation. The advantages of BNE were evident in the scenarios with high densities, compared to those with low ones.

## 5 Discussion and Conclusions

This paper evaluated evacuation models that incorporate game theory (i.e. Bayesian Nash Equilibrium) within multi-agent systems with the aim of providing more realistic simulations of the movements and behaviours of evacuating pedestrians. Pedestrian agents adopting BNE were found to be more representative than agents with other behavioural patterns (SR and RF). At each step, they predict the next move of their neighbours and then avoid the most congested patches on their way to the exits, resulting in a relatively high comfort level. It was hypothesised that agents adopting BNE may provide a more forward-looking and representative behavioural model for pedestrian evacuation, as BNE agents sought to avoid congestion instead of directly moving to the exits or blindly following others.

A series of simulation experiments were undertaken to evaluate the role of BNE in pedestrian evacuation. The results demonstrate the positive impacts of BNE on reducing evacuation time and improving individual comfort during pedestrian evacuation, which was consistent regardless of the number (density) of agents. Agents with BNE displayed more efficient and intelligent behaviours during simulation compared with RF and SR agents, suggesting how simulation models that incorporate such behaviours for pedestrians during evacuation could be better represent real world scenarios. The individual decision-making process based on BNE is easily adaptable to other pedestrian simulations relating to flooding or fire and has the potential to fill the gap of a lack of forward-looking, intelligent individual behavioural model in ABMs for pedestrian evacuation.

The BNE agents in this model were assumed to be able to independently determine their next actions after considering environmental factors, as well as probabilities of their neighbours' movement and decisions, which in turn, affect the subsequent steps of other agents. That is, agent decisions and movement with BNE varied based on their interactions with surroundings, predictions of neighbour's actions and their own decisions, making the model outputs more realistic. This is different to previous simulation models of pedestrian flow (e.g. Jiang et al. 2010; Teknomo 2016; Lu et al. 2017). Thus far, research incorporating game theory and agent-based modelling has mainly focused on comparing game theory and ABM approaches (Norri et al. 2021) and the combination of ABMs and simple game theory (Levy et al. 2018). Studies of cooperation under Bayesian game theory and pedestrian evacuation have generally focused on individual decision-making over exit choice rather than pedestrian movements during evacuation process (Mesmer and Bloebaum 2014). The models described in this paper address this gap by incorporating complex game theory within an ABM approach at the individual agent level. Rather than simulation of exit selection, this model simulated pedestrian decisions using BNE for each time step, which more closely matches the reality of people avoiding crowd spaces in evacuation with routes that might be not the shortest path. Under BNE, the expected comfort utility of patches was constantly being updated by the varying distributions of other nearby agents at each time step. This allowed BNE agents to predict the next move of other agents and then to avoid the most congested areas during evacuation, in contrast to SR agents who move directly to the exit causing large congestions and resulting in longer evacuation time, and RF agents who randomly follow a neighbour resulting in smaller groups gathering but still slower evacuation times. The order of the effects of three behavioural patterns from high to low level is: BNE, RF, SR.

The proposed BNE model has a number of limitations: 1) Attributes such as moving speed and comfort utility need to be further calibrated to match evacuation movements in the real word; 2) Further simulation experiments need to be conducted with a wider range of parameter configurations, and different static variables (e.g. exit size) to further understand the effects of BNE on evacuation process from various perspectives; 3) Real or other existing evacuation models datasets are needed to validate this model; 4) The role of space played in decision-making process could be further explored by adding blockades into the simulation space before or during evacuation; 5) Greater self-organizing behaviours among pedestrians (e.g. competitive behaviour) need to be considered to make the model easily adaptable to a broader range of pedestrian studies. Such research is necessary to determine the full potential of the use of Bayesian Nash Equilibrium in pedestrian decision making in evacuations and to further advance the application of Bayesian game theory in agent-based pedestrian modelling.

# Appendices

**Appendix A**: Facets: The variations of evacuation time with different number of pedestrians (Pattern: BNE mixed with RF)

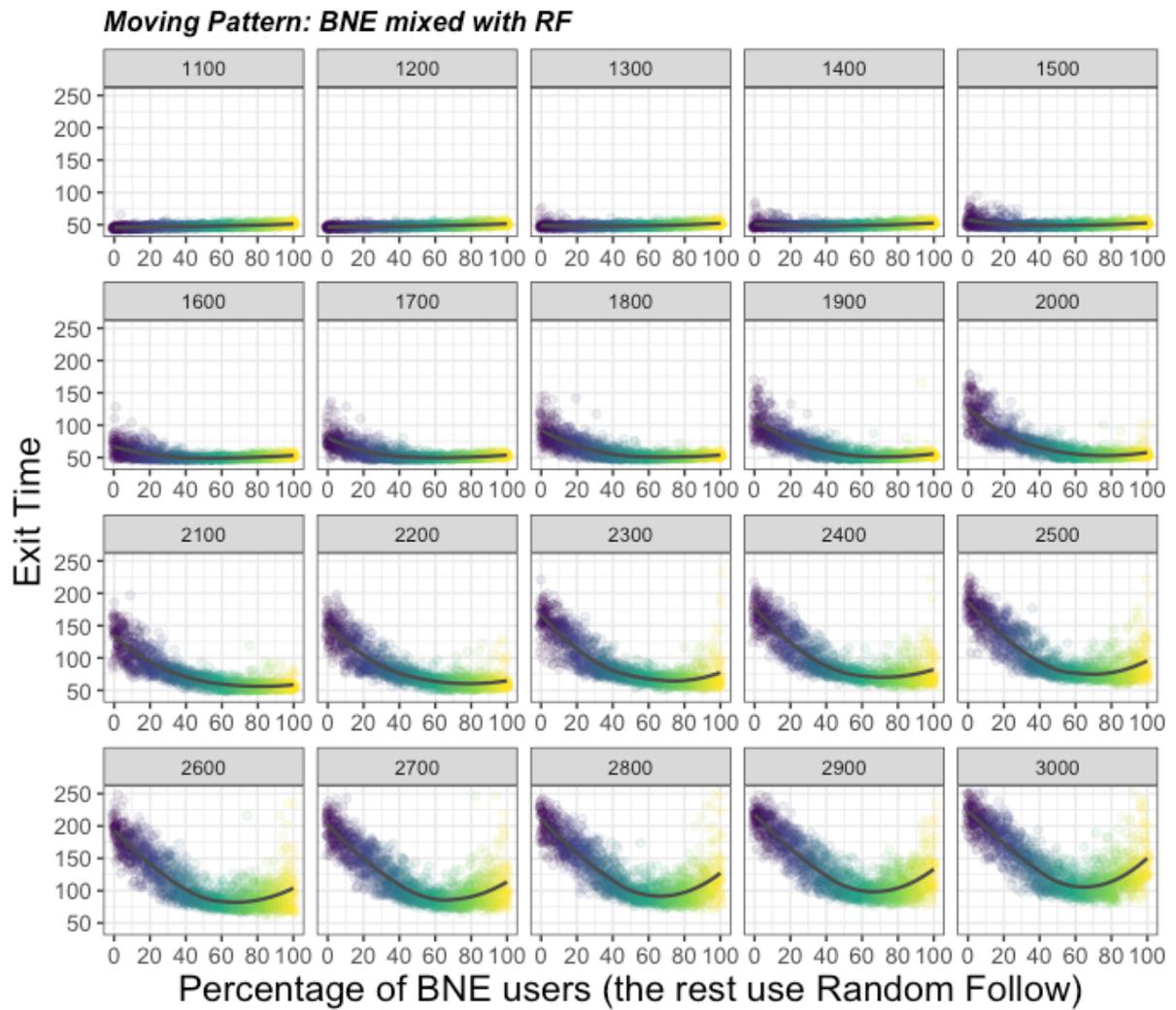

**Appendix B**: Facets: The variations of evacuation time with different number of pedestrians (Pattern: BNE mixed with SR)

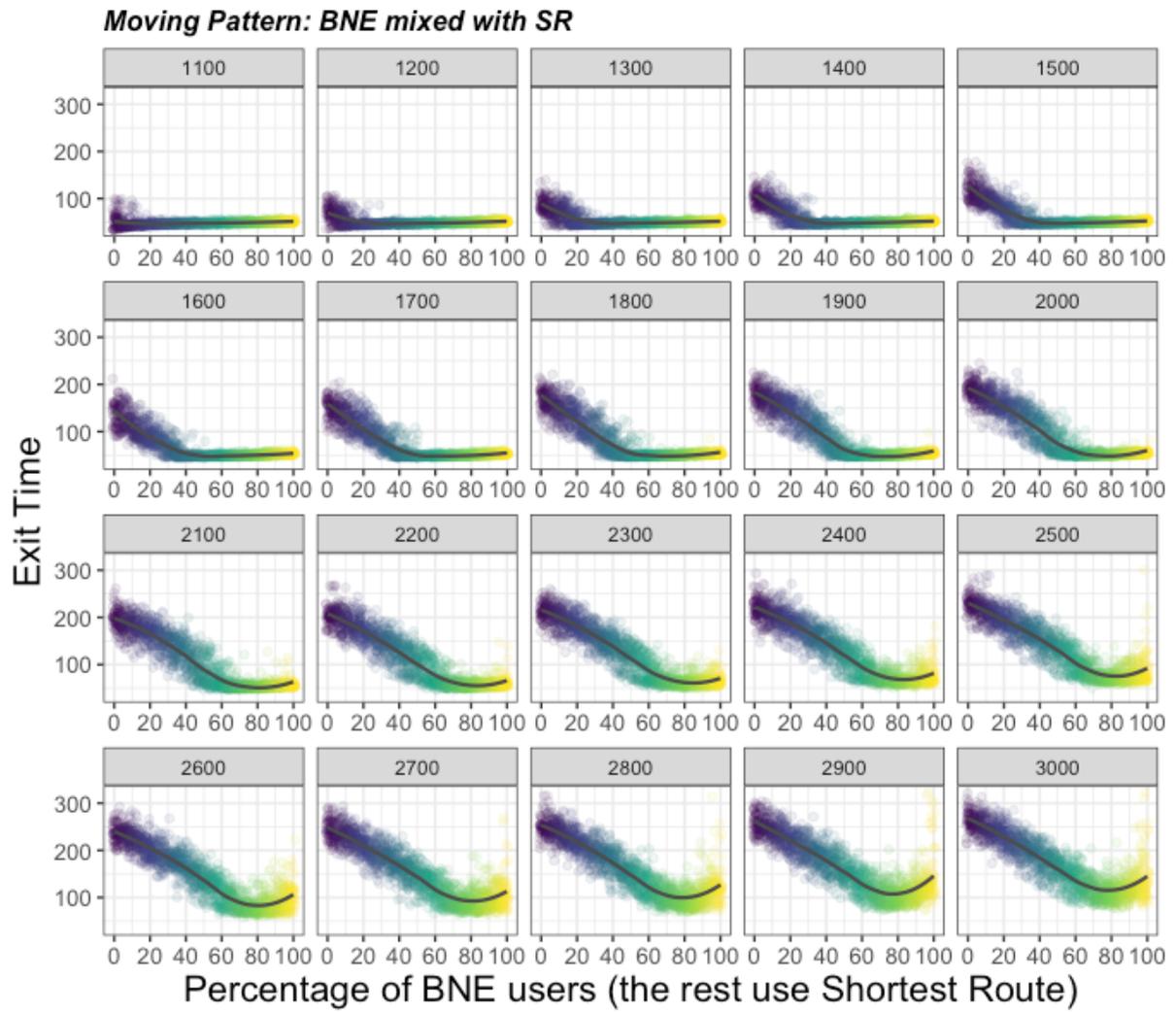

**Appendix C**: Facets: The variations of average expected comfort utility with different number of pedestrians (Pattern: BNE mixed with RF)

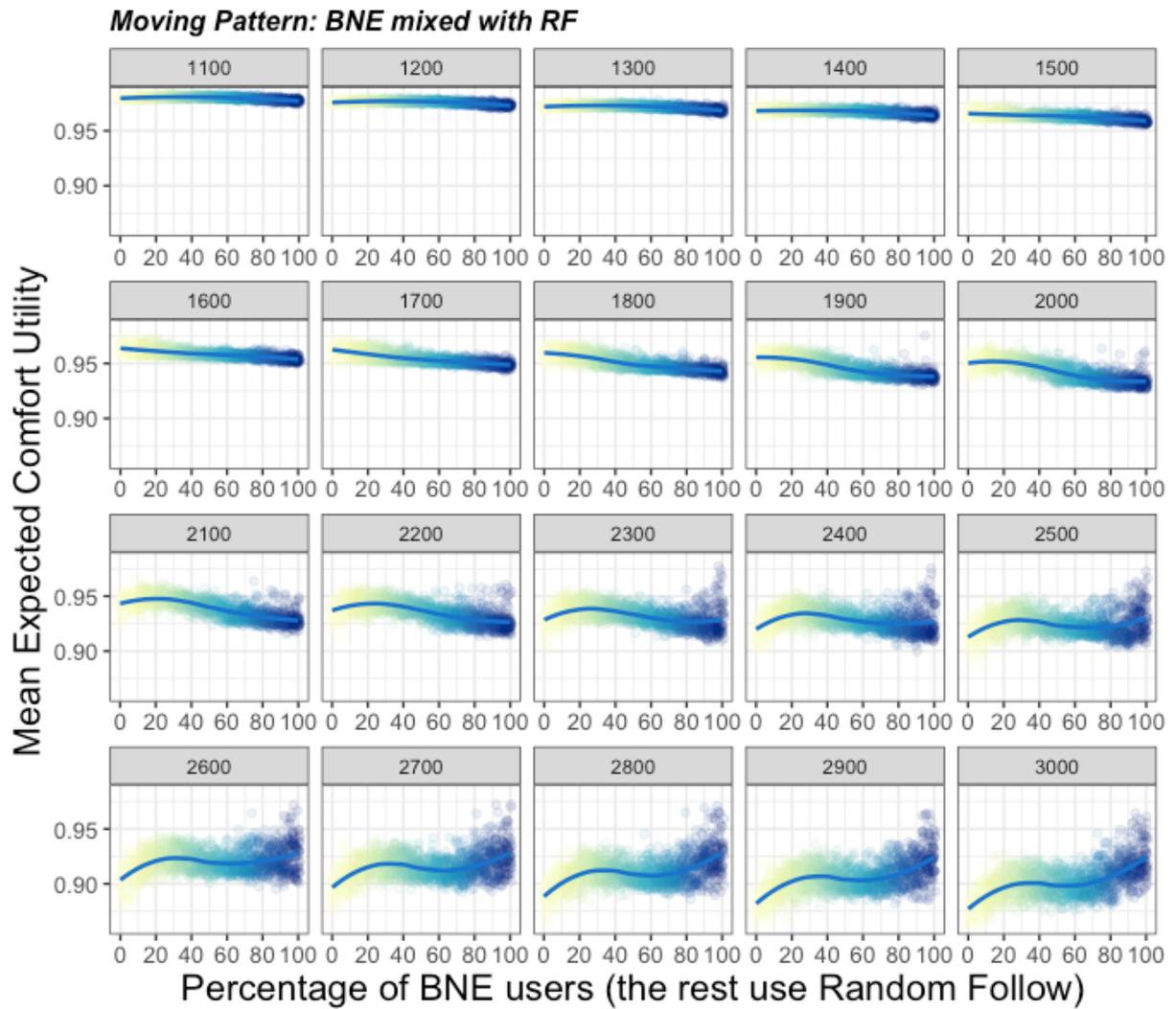

**Appendix D**: Facets: The variations of average expected comfort utility with different number of pedestrians (Pattern: BNE mixed with SR)

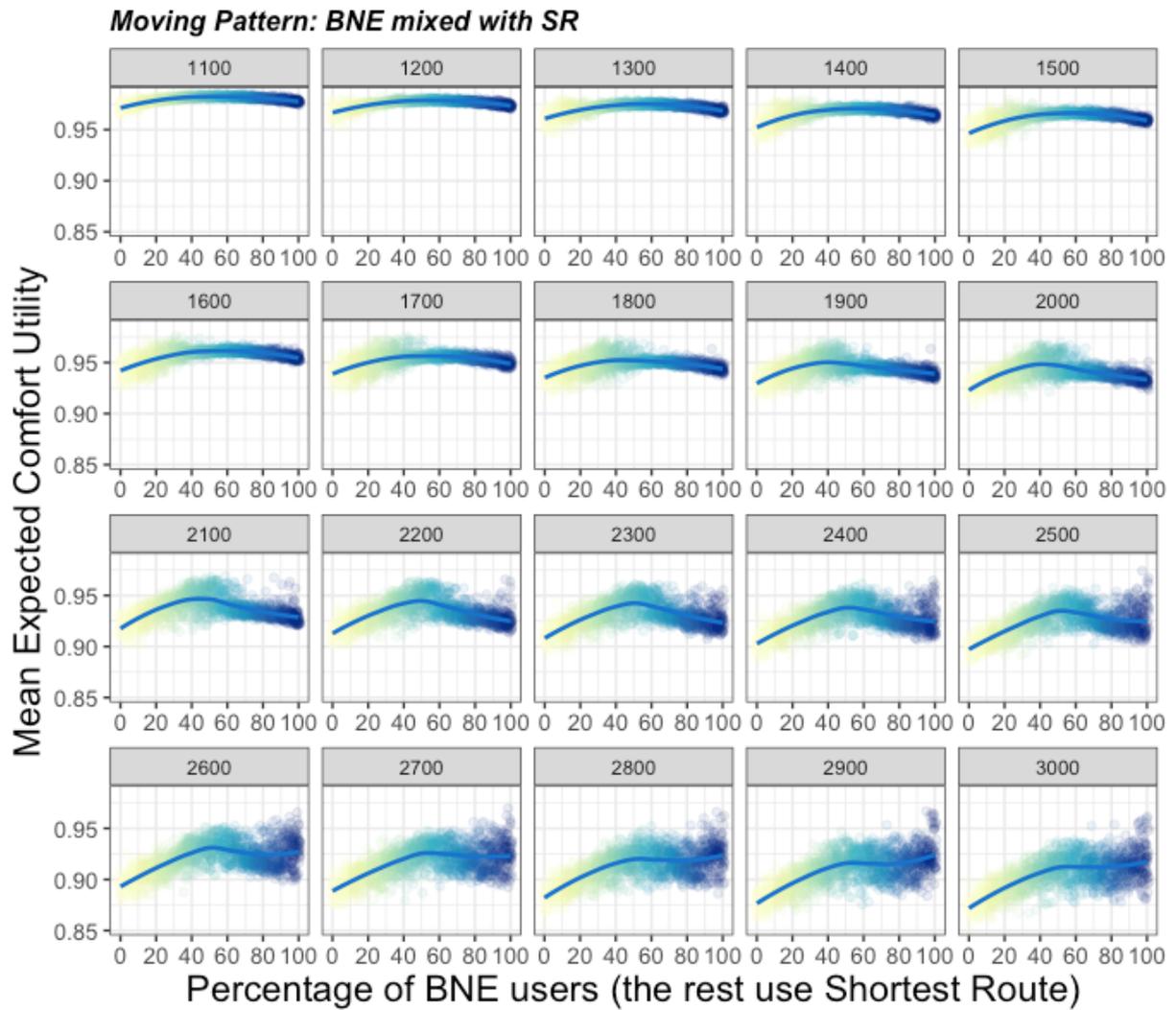